# CASCADE MIXING, A NEW KIND OF PARTICLE MIXING PHENOMENON

**B. Kayser**

Division of Physics, National Science Foundation

4201 Wilson Blvd., Arlington VA 22230, USA

**L. Stodolsky**

Max-Planck- Institut für Physik

Föhringer Ring 6, 80805 München, Germany

**ABSTRACT**

We discuss "cascade mixing", where one particle mixture, say a $B^0$, leads to another, say a $K^0$. A simple analysis is possible in the amplitude approach, which avoids "collapses of the wavefunction" and is explicitly covariant. Some novel possibilities, both of conceptual and perhaps of experimental interest, arise. For example, we explain how such processes can allow one to "tune", in principle, the phase relations in a particle mixture. Also, effects arise involving combinations of the mass differences of two particle mixtures. We explain how an intermediate measurement may play the role of a regenerator, so that in principle regeneration-like effects can be induced for the $B^0$ and $D^0$ systems, despite their short flight paths. The analysis of such process with respect to CP and the distinction between "direct" and "indirect" CP violation is discussed.

Particle mixing, as typified by $K^0$ meson phenomena, has long been a paradigm of fascinating quantum mechanical behavior. It has served as a model for many interesting systems and questions, and in addition has permitted very precise studies of $K^0$ mesons involving the fundamental symmetries. In this note, we would like to consider a new set of related phenomena of this type, again a striking example of quantum mechanical behavior, and perhaps also useful in studying properties of the particles involved.

This new class of effects arises because of the possibility, which did not exist for the classic $K^0$ system by itself, of one type of mixing system, for example the $B^0$ turning into another, say the $K^0$.

Such processes may seem at first rather subtle to analyze, which may be one reason they have not been extensively discussed. One must deal with the effects of measurements of interfering systems at different space-time points, and if one adopts the traditional language of "the collapse of the wavefunction" the problem appears rather confused and complicated. One of the questions, for example, has to do with how a "measurement" of just part of a system affects its further evolution. In addition, some of the cases involve relativistic effects, like $B^0 \to K^0$ where the $K^0$ is fast in the $B^0$ rest frame, so that questions concerning in which lorentz frame to perform the "wavefunction collapse" can no longer be ignored or answered by simple intuition. However, in a recent Letter[1] devoted to understanding EPR experiments without "the collapse of the wavefunction", we presented a simple framework for dealing with such problems: the "amplitude approach". This approach does not invoke the awkward wavefunction "collapses", is explicitly covariant and transparent physically. This framework, as we briefly mentioned in ref [1], makes it possible to handle the "cascade mixing" problem where one mixing system turns into another.

The fundamental point in ref [1] was to avoid use of the wavefunction and to focus on the amplitude. This permits a simple and intrinsically covariant description, without "collapses". The basic dynamical recognition is that the phase which plays the principle role in mixing is simply mass($m$) times proper time ($\tau$). The only subtle point is that $\tau$ must be correctly understood, namely as the proper time connecting two space-time points, without any reference to a particular particle.

The simplest situations, which we begin with, are those that can arise in a single beam, which we call the "single-arm case". We suppose, however, that there are two detectors (at least) viewing this single beam. We imagine that in this beam a process like $B^0 \to K^0$ with certain decay products can take place.

After the single-armed case, one may consider the "two-armed" case where cascade processes take place in EPR-like configurations like $\phi \to K^0 \; K^0$, $\psi(3770) \to D^0 \; D^0$ or $\Upsilon(4S) \to B^0 B^0$. We touch on this briefly.

As in previous work, we confine ourselves to two-state systems like $K^0$, $B^0$ and $D^0$. More complicated cases, as might arise for neutrinos where we have three species, can be dealt with by the same methods.

## I. SINGLE-ARMED PROBLEMS

We recall that in the amplitude method we must studiously avoid talking about which particles one "really had" in intermediate states. These constitute the "interfering alternatives"[2], and one should not try to determine them any more than one tries to determine through which slit the photon goes in the two-slit experiment. On the other hand, it is important to include the production and detection processes in the description.

To make the problem clear, we present a definite example. Let a fixed production process $P$ create a $B^0$ meson. Then let the $B^0$ decay via $B^0 \to J/\psi K^0$, and finally let the $K^0$ decay to two charged pions. In these considerations one presumes that localized detection at well defined space-time points is valid[1], so that in addition to specifying the decay channels, we also need the space-time points of the various occurrences. Taking the production process at the origin (0), we then call (1) the spacetime point of $B^0 \to J/\psi K^0$ and (2) the spacetime point of $K^0 \to \pi^+\pi^-$. In the quasi-classical limit in which we work we have $x \approx vt$ from classical kinematics. Since there are two possible $K^0$ states and two possible $B^0$ states, there are four possible interfering alternatives leading from the initial to the final state. Let the amplitude for a process involving a $B^0$ or $K^0$ of a given type (like "long" or "short") be denoted with an index n or m:

$$A(P \to B^0_m) = A_m(P \to B^0),$$

so that this amplitude has the appearance of a "spinor" in the two-dimensional $B^0$ space. Similarly, we have the amplitudes $A_m(K^0 \to \pi^+\pi^-)$, again a "spinor", and finally $A_{nm}(B^0 \to J/\psi K^0)$, a matrix.

Now according to ref [1], the total amplitude for the whole process is

$$A_n(P \to B^0) \, e^{-im_B^n \tau^n(1,0)} \, A_{nm}(B^0 \to J/\psi K^0) \, e^{-im_K^m \tau^m(2,1)} \, A_m(K^0 \to \pi^+\pi^-) \,, \quad (1)$$

where a sum over n and m is intended, giving the total amplitude as the sum of the four contributions. The first $\tau$ is the proper time between the production and the detection (1), and the second $\tau$ is the proper time between the detection (1) and the detection (2). The indices $m, n$ refer to a given physical mass eigenstate like "long "(l) or "short"(s) for $K^0$, and their analogues for $B^0$ and $D^0$. As discussed in ref [1], if there is a lorentz frame where the two mixing particles are both non-relativistic then we may work in this frame and safely ignore the index on $\tau$. On the other hand if the relativistic $\gamma$ factors of the mixing particles differ widely, as might be the case for neutrinos , then it is necessary to take account of these indices. The masses are understood to have an imaginary part to account for the width or lifetime of the particles .

We stress that it is important to think of a detection as taking place at a given space-time point, independent of which mass eigenstate is involved. That is, in each of the four contributions to the amplitude the detections always take place at the same spacetime points. Otherwise one may get into difficulties[3]. In the present discussion, the detection is always a decay process, although in principle it can also be something else such as a scattering .

We can write Eq (1) in a compact matrix notation as

$$A(0)S(1,0)A(1)S(2,1)A(2) \,, \quad (2)$$

The $A$ are the various production and detection amplitudes while $S(1,2)$ is the propagation phase- and- decay matrix $e^{-im_K \tau(2,1)}$, constructed from the 2x2 mass matrix $m_k$, and similarly for $S(1,0)$ . $S$ is a two-by-two matrix in the space of the $K^0$ (or $B^0$ ) states. Because of the imaginary parts $\Gamma$ of the masses $m_K$, the mass matrix is not hermitian and so $S$ is not necessarily unitary or proportional to a unitary matrix. Furthermore its eigenvectors may not be orthogonal. This may be dealt with by introducing the "duals" $|s^d>, |l^d>$ to the two non-orthogonal mass eigenvectors (like "long" and "short") $|s>, |l>$ so that $<s^d|s>=1, <s^d|l>=0$ and so forth[4]. We then have $S(2,1) = (e^{-im_K^s \tau^s(2,1)})|s><s^d| + (e^{-im_K^l \tau^2(2,1)})|l><l^d|$, which it may be verified, propagates the two mass eigenvectors suitably. Although in

our examples the superscripts on the $\tau$'s may usually be dispensed with, we have left them on to indicate the general case.

Since $A(0)$ and $A(2)$ are "spinors" with respect to transformations in the 2x2 mixing space while $S$ and $A(1)$ are matrices, Eq (2) may be viewed as the element of a certain matrix $M = S(1,0)A(1)S(2,1)$ between the "spinors" $A^*(0)$ and $A(2)$. Squaring to obtain the rate gives

$$Rate \sim Tr[\rho^P M \rho^D M^\dagger], \tag{3}$$

where $\rho^P = A^*(0)A^*(0)^\dagger$ is a production "density matrix" and $\rho^D = A(2)A(2)^\dagger$ is a detection "density matrix". The detection at (1) appears on a somewhat different footing than that at (2) since the detection at (1) involves two indices, i.e. is a matrix in the 2x2 space. If $\rho^P$ and $M$ commute this becomes

$$Rate \sim Tr[\rho^P \rho^D M^\dagger M], \tag{4}$$

while if $\rho^D$ and $M$ commute we have

$$Rate \sim Tr[\rho^P \rho^D M M^\dagger], \tag{5}$$

## II. PROPERTIES OF THE FACTORS

It is useful to consider the properties of the factors in Eq (3) under various assumptions. In particular we consider the influence of the assumptions of CP and CPT conservation, as well as the choice of detection and production channels. It will be convenient, due to the flavor selection rules of the standard model, to work in a specific basis in the 2x2 space, the flavor basis. With the usual pauli matrices this is the basis where $\sigma_3 |K^0> = +|K^0>$ and $\sigma_3 |\bar{K}^0> = -|\bar{K}^0>$, and similarly for the definite flavor states of $B^0$ and $D^0$.

*Properties of M:* We have $M = S(1,0)A(1)S(2,1)$ For the propagation matrices $S$, CPT invariance means that there is no $\sigma_3$ term in the mass matrix, while CP invariance means that the antisymmetric $\sigma_2$ term is absent. Hence in the limit of good CP and CPT, $S = e^{-i(m_0 + m\sigma_1)\tau}$, where $m_0$ is the average mass of the two neutral particles and $m$ is half the mass difference. Since the $m_0$ term is proportional to the identity matrix and commutes with everything, we will drop it in most of the following.

For the detection amplitudes there are two important cases: detection of a definite flavor, as via the channel $K^+ e^- \bar{\nu}$, and detection via a self-conjugate system, where in the limit of CP conservation a definite CP can be assigned to the detection process. This latter case can then be further divided into the cases of CP even and CP odd detection. In the CP good limit, the even CP of the $J/\psi$ and the necessity of $l = 1$ in the transition give a transition of odd CP. Therefore we have, in our flavor basis, $A(1) \sim \sigma_3$. On the other hand with an s-wave pion pair, like $\pi^0 \pi^0$, we have CP even detection: $A(1) \sim I$. Hence, in the limit of good CP and CPT, we have for CP even detection at (1), $A(1) \sim I$,

$$M \sim e^{-i\sigma_1[m_B \tau(1,0) + m_K \tau(2,1)]} \tag{6}$$

and for CP odd detection at (1) with $A(1) \sim \sigma_3$, and so

$$M \sim \sigma_3 e^{-i\sigma_1[-m_B \tau(1,0) + m_K \tau(2,1)]} \tag{7}$$

when the anticommutation properties of the $\sigma$ are used.

For a non-self-conjugate detection like $K^+ e^- \bar{\nu}$, flavor considerations alone give $A(1) \sim I \pm \sigma_3$, which results in the sum of Eq (6) and Eq (7). CP conservation would then fix the amplitude for the conjugate process $K^- e^+ \nu$ at (1).

*Properties of the $\rho$:* We now turn to the $\rho$.

For $\rho^D$ we have again the three main cases of flavor, CP even, and CP odd detection. For example when our second mixing system is $K^0$ these could correspond to the channels $\pi^+ e^- \bar{\nu}$, $\pi^+ \pi^-$ or $\pi^0 \pi^0 \pi^0$. For a given final state $f$, $[\rho^D]_{ij} = A(i \to f) A^*(j \to f)$, which in the flavor basis leads to

$$\rho^D = (I \pm \sigma_3), (I + \sigma_1), (I - \sigma_1) \tag{8}$$

for the three cases respectively, always in the CP conserving limit.

For $\rho^P$ the two main cases would appear to be the flavor tag, $\rho^P = (I \pm \sigma_3)$, or the CP tag $\rho^P = (I \pm \sigma_1)$.

*Sums or Mixed States:* So far we have dealt only with pure states or amplitudes. A single amplitude, corresponding to a single production mechanism and single detection mechanism was assumed. If we now suppose a sum over different production or detection channels, the $\rho$ will become sums, like for the usual density matrix.

For $\rho^D$ we may observe that when *all* final states are summed over, we have essentially the width or $\Gamma$ term in the mass matrix[5] of the second mixing system, so the same considerations apply as for the mass matrix, namely with CP and CPT good,

$$\rho^D \sim \Gamma \sim (I) + (\sigma_1), \tag{9}$$

where no particular proportions between the components is implied. For $K^0$ there is a large $(\sigma_1)$ term reflecting the large lifetime difference between the mass eigenstates, while for the heavier analogues this term tends to be relatively small. With account of CP violation there is also a small $\sigma_2$ component, reflecting "direct" CP violation.

For $\rho^P$ the most typical inclusive sum would be for an untagged production mechanisms where both flavors are produced incoherently and equally, so that $\rho^P \sim I$. Naturally, partial sums of various kinds will be more complicated and must be examined individually.

Finally, there is an implicit label on the $M$ refering to the detection channel at (1). If the matrix amplitude $A(1)$ may be commuted through the propagation factors $S$ so that the expression $A(1)A^\dagger(1)$ appears in $MM^\dagger$, then the detection at (1) may be handled by defining a $\rho$ which is a sum over products of matrix pairs.

### III. SOME FEATURES

Because of the many amplitudes and parameters involved, it will require an extensive analysis to sort out the many different cases and possibilities; not to speak of understanding the experimental limitations. However we would like to draw attention to some of the amusing new possibilities which suggest themselves.

*Tuning the mixture:* One is the possibility, which now in principle exists, of "tuning" the state of a particle mixture. One may read Eq (2) to say that after the flight path (1,0) a $B^0$ mixture arrives at the detection $A(1)$ with some set of phase-and-magnitude relations. After the detection it leaves the point (1) as a certain $K^0$ mixture, which after a further flight path is detected in a certain way at (2).

Now the location of the point (1) can be varied, in principle. Due to the mass difference in the $B^0$ system, this induces a continuous variation of the "incoming" $B^0$ mixture at (1) and thus an adjustable $K^0$ particle mixture is "outgoing" from (1). In the traditional language of $K^0$ physics the detection $A(1)$ plays the role, in a sense, of a regenerator, a piece of physical material; we might say we have "detection regeneration".

We thus have a method, at least in principle, of producing a continuously adjustable $K^0$ mixture outgoing from point (1). In the past a certain degree of adjustment of the parameters of a $K^0$ particle mixture was possible by arranging for suitable regenerators and adjusting the flight paths in the beam. Our "cascade mixing" however, allows a different approach, one which can also be applied to $B^0$ and $D^0$ as well.

*Mass-Difference Differences:* Relations Eq (7) and Eq (6) are intriguing because they suggest oscillation effects where the differences or sums of mass differences of the two mixing systems might appear.

To manifest such effects, however, the processes must be correctly chosen. If the rate expression Eq (3) leads to simply an exponential times its complex conjugate then oscillation effects will be absent. Hence we wish to avoid arriving at one of the two forms Eq (4) or Eq (5); i.e. $M$ should not commute with either of the two $\rho$'s. This indicates using the flavor tag production $\rho^P = (I \pm \sigma_3)$ and flavor detection $\rho^D = (I \pm \sigma_3)$. This is like classic $K^0$ experiments where a definite strangeness is produced, as tagged by a hyperon, and then strangeness oscillations are studied in the further decays or interactions. Indeed, one verifies that with these choices and taking our example of CP odd detection at (1), that is with Eq (7), that the rate is proportional to

$$e^{-\Gamma_{B_2}\tau(1,0)-\Gamma_{K_1}\tau(2,1)} + e^{-\Gamma_{B_1}\tau(1,0)-\Gamma_{K_2}\tau(2,1)}$$

$$+2e^{-\Gamma_B\tau(1,0)-\Gamma_K\tau(2,1)}cos[-(m_B + m_B^*)\tau(1,0) + (m_K + m_K^*)\tau(2,1)] \quad (10)$$

where $\Gamma_B$ is the average lifetime of the two $B^0$ states, $\Gamma_K$ that of the $K^0$ states, $\Gamma_{B_1}$

that of the CP even $B^0$ eigenstate, $\Gamma_{B_2}$ that of the CP odd $B^0$ eigenstate, and so forth. If we had used CP even detection at (1), that is Eq (6), we would have

$$e^{-\Gamma_{B_1}\tau(10)-\Gamma_{K_1}\tau(2,1)} + e^{-\Gamma_{B_2}\tau(1,0)-\Gamma_{K_2}\tau(2,1)}$$

$$+2e^{-\Gamma_B\tau(1,0)-\Gamma_K\tau(2,1)}cos[+(m_B+m_B^*)\tau(1,0)+(m_K+m_K^*)\tau(2,1)] \quad (11)$$

where in the oscillatory term the sign of $m_B$ is now reversed. (Flavor detection at (1) then gives a combination of both). Recall that $m$ represents half the mass difference so the oscillations are simply at the frequency corresponding to the mass difference itself.

We thus arrive at expressions with the amusing feature that they show oscillations involving both sets of mixing masses together. Therefore there is, at least in principle, the possibility of comparing various mass differences through their effects in one physical system . For example, holding the sum of the $\tau's$ fixed while varying their difference creates an effect involving the difference of the differences. Quantitatively, both components of the argument of the cosine are of about the same size. That is, the lifetime times the mass difference are roughly the same for $K^0$ and $B^0$ . Since the bulk of events will occur when a $\tau$ on the order of the lifetime, both terms are about of equal importance. This will of course be different for $D^0$ where the mass difference is small compared to the inverse lifetime. We should perhaps stress that we use the $B^0$ system and $B^0 \to J/\psi K^0$ merely as an example. For the discussion of the CP conserving limit this may be somewhat artificial since the $B^0$ are expected not to be a good CP eigenstates and substantial CP violation is hoped for in $B^0 \to J/\psi K^0$ .

*CP Test:* Note that the form of Eqs (10,11) only depend on whether the detection at (1) is CP odd or even. First of all, this means that all processes of a given CP type may be added together, helping in the collection of a large number of events to study the oscillations. Secondly, this may be used as a CP test since, evidently, observation of both types of behavior in one process, that is to say oscillations corresponding to neither the sum nor the difference of the masses, or decay patterns not corresponding to CP conservation or CP flip at (1), but rather a combination of both, would indicate CP violation in the amplitude $A(1)$. This might be expressed in another, perhaps experimentally more striking way, by saying if CP is good, the ratio of any two processes with the same CP type detection at (1) must be constant as the $\tau$'s are varied.

Observe that this kind of CP violation if seen, would necessarily indicate "direct" CP violation, that is violation in a decay amplitude, as opposed to a mass-mixing, "superweak"[6] type of CP violation. Consider the comparison of two processes when the only CP violation in the problem is due to "mass mixing"; the $A(1)$ for the two processes are proportional, and so the overall $\tau$ behavior is the same for both. On the other hand, with "direct" violation in the decay at (1) we will have different combinations of $A(1)$ in general and hence different behavior from one process to another.

## IV. DOUBLE-ARMED CONFIGURATIONS

In the "amplitude approach" the double-armed configuration, by which we mean configurations that begin like $\Upsilon \to B^0 \, B^0$ is not much different than the single-armed situation. The only additional subtleties may arise from a (anti)symmetrization which may be necessary in adding interfering alternatives. For example, since $\Upsilon \to B^0 \, B^0$ is an $l = 1$ decay, we have $A(\Upsilon \to B_H B_H) = A(\Upsilon \to B_L B_L) = 0$ and $A(\Upsilon \to B_H B_L) = -A(\Upsilon \to B_L B_H)$. $H, L$ refer to the two mass eigenstates "heavy" and "light" and the first and second $B$ refer to different momentum states, i.e., different directions in space. Let $X$ refer to the complete array of parameters and detectors specifying the right arm and $Y$ that for the left arm. The specification of the space-time points for the various detections is meant to be included. Then we will have, according to whether there is a symmetrization or an antisymmetrization

$$A(B_L \to X)A(B_H \to Y) \pm A(B_L \to Y)A(B_H \to X) \qquad (12)$$

where for the $l = 1$ decays $\Upsilon \to B^0 \, B^0$, $\psi \to D^0 \, D^0$ and $\phi \to K^0 \, K^0$, we have an antisymmetrization.

We can divide the two-armed configuration into two major cases: single cascades and double cascades. In the first case, on one side, say the $X$ side, we have just one mixing system while on the $Y$ side we have a "cascade"; in the second case we have a "cascade" on the $X$ side also. In $\Upsilon \to B^0 \, B^0$, an example of the first case would be $B^0 \to D^+ X^-$ on the $X$ side while on the $Y$ side there is some process with $B^0 \to D^0$, while in the second case there is $B^0 \to K^0$ on both sides.

The two cases differ in how far it is necessary to pursue the sequence of events on the $X$ side. This is the issue of where we draw the line between the "experiment" and the "observer". As discussed in ref [1], this line may be drawn where there is no danger of further interfering alternatives[7]. Although in the first case the $D^+X^-$ system will generally further decay or interact, we can nevertheless stop the analysis at this point. This is because any given further state on the $X$ side will be clearly attributable to $D^+X^-$, so we may as well stop here; any further steps will be prefaced with a common factor $A(B \to D^+X^-)$. On the other hand, with a cascade on the $X$ side there are further interfering amplitudes and it is necessary to pursue the process further.

## V. CP IN THE DOUBLE ARMED CONFIGURATION

We briefly mention some of the important points concerning the double armed configuration with respect to CP; most of these are well known[8],[9,10] but it is perhaps useful to repeat them in the present context. First of all, starting with $\phi$, $\psi$ or $\Upsilon$, and if CP is conserved and only CP eigenstates are detected, then the product of all detections must be odd in the sense of CP. For $\phi$, for example, if pion pairs are observed on both sides, giving an even CP overall, this cannot correspond to $K_1, K_2$ and CP conserving decays.

In particular this means that with CP conservation and detection of CP eigenstates, the configuration with the same particles on both sides is forbidden, since the total CP would be automatically even. Hence observation of this configuration implies CP violation. In practice, if $K^0$'s are involved CP violation at the $10^{-3}$ level is of course expected; effects at substantially more than this level would indicate new sources of CP violation.

Allowing for CP violation, the same particles may occur on both sides in general, but because of Eq (12), the total amplitude must still vanish when $X = Y$ in the sense that the space-time specification is included, that is when the various proper times are the same on both sides. On the other hand it suffices for just one of the $\tau$ to differ to obviate the cancellation of the two parts of Eq (12).

A final point, concerning the case where CP is violated but $X \neq Y$, brings us back to the historical origins of the EPR-like idea[10] where the study of $\phi \to K^0$ $K^0 \to (\pi^+\pi^-)(\pi^0\pi^0)$ was proposed. Eq (12) with the minus sign resembles the determinant of a 2x2 matrix. Since a determinant is zero if the columns are linearly dependent, it thus gives zero if the $B_L$ and $B_H$ amplitudes are simply proportional to each other. Now in a pure mass mixing or "indirect" (superweak) model of CP violation the decay amplitudes are indeed proportional to each other, since everything goes by way of a common state. However, the amplitudes in Eq (12) also involve propagation factors in addition to decay amplitudes. But their effect can be eliminated by choosing a symmetric configuration with the $\tau(1,0)$ the same on both sides.

Hence for the CP violating decay into states of the same CP on both sides, Eq (12) vanishes in this symmetric configuration for *any* pair of final states unless there is a "direct" ($\epsilon'$-like) contribution to a decay amplitude. This remains true for "cascades"; if the $\tau(1,0)$ are the same on both sides, then Eq (12) with the minus sign vanishes for any such pair of final configurations unless there is a "direct" CP violation.